\documentclass[10pt,twocolumn]{IEEEtran}
\usepackage{graphicx}
\usepackage{float}
\usepackage{algorithm}
\usepackage{booktabs,subcaption,amsfonts,dcolumn}
\usepackage{amstext}
\usepackage{amsmath}
\usepackage{amsfonts,amssymb}
\usepackage{amsmath,tabu}
\usepackage{amssymb}
\usepackage[dvips]{epsfig}
\usepackage{epsfig}
\usepackage[dvips]{graphicx}
\usepackage{hyperref}
\usepackage{multirow}
\usepackage{multirow} 
\usepackage[utf8]{inputenc}
\usepackage[english]{babel}
\usepackage[labelsep=period]{caption}
\usepackage{subcaption}
\usepackage{float}
\usepackage{ltablex}
\usepackage{cite}
\usepackage{lipsum}
\usepackage{flushend}
\usepackage{algpseudocode}
\usepackage{setspace}

\usepackage[justification=centering]{caption}

\def\b1{{\boldsymbol{1}}}
\def\c1{{\textcircled{a}}}

\def\bu{{\mathbf{u}}}
\def\bv{{\mathbf{v}}}

\def\bC{{\mathbf{C}}}

\usepackage{cleveref}
\usepackage{xcolor}
\usepackage{soul}

\usepackage[english]{babel}

\title{\huge Pearson-Matthews correlation coefficients for binary and multinary classification and hypothesis testing
\author{Petre Stoica and Prabhu Babu
\thanks{Petre Stoica is with the Division of Systems and Control, Department of Information Technology, Uppsala University, Uppsala, Sweden 75237 and Prabhu Babu is with the Centre for Applied Research in Electronics, Indian Institute of Technology, Delhi 110016, India (email: ps@it.uu.se, Prabhu.Babu@care.iitd.ac.in).
Petre Stoica's work was supported in part by the Swedish Research Council (VR grants 2017-04610,  2016-06079, and 2021-05022).}}
}

\begin{document}
\maketitle
\begin{abstract}
The Pearson-Matthews correlation coefficient (usually abbreviated MCC) is considered to be one of the most useful metrics for the performance of a binary classification or hypothesis testing method (for the sake of conciseness we will use the classification terminology throughout, but the concepts and methods discussed in the paper apply verbatim to hypothesis testing as well). For multinary classification tasks (with more than two classes) the existing extension of MCC, commonly called the $\text{R}_{\text{K}}$ metric, has also been successfully used in many applications. The present paper begins with an introductory discussion on certain aspects of MCC. Then we go on to discuss the topic of multinary classification that is the main focus of this paper and which, despite its practical and theoretical importance, appears to be less developed than the topic of binary classification. Our discussion of the $\text{R}_{\text{K}}$ is followed by the introduction of two other metrics for multinary classification derived from the multivariate Pearson correlation (MPC) coefficients. We show that both $\text{R}_{\text{K}}$ and the MPC metrics suffer from the problem of not decisively indicating poor classification results when they should, and introduce three new enhanced metrics that do not suffer from this problem. We also present an additional new metric for multinary classification which can be viewed as a direct extension of MCC.  
\end{abstract}
\section{introduction}
We consider a situation in which a supervised classification experiment for $K$ classes has produced a confusion matrix (aka error or frequency table) whose element $C_{kl}$ denotes the number of times the $k-$th class has been classified as the $l-$th class ($k,l=1,\cdots,K$). In the case of two classes (i.e. $K=2$), the entries of the confusion matrix bear specific names, see Fig. 1.

The names of $\{C_{kl}\}^2_{k,l=1}$ in Fig. 1 are predominantly used in statistics, chemistry, bio-medicine, bio-informatics, and related applications. In radar and a couple of other engineering applications the corresponding names are: $C_{11}$ = detection, $C_{12}$ = miss, $C_{21}$ = false alarm, and $C_{22}$ = rejection.

A common problem in the above applications and a few others, such as machine learning and image analysis, is combining the numbers in the confusion matrix into a scalar metric (or score) that can be used for instance to compare the performance of different classifiers. A large number of metrics have been suggested for this purpose (see, e.g., \cite{chicco}, \cite{grand}, \cite{ref6,ref7,ref8,ref9},\cite{chicco2,jurman} and \cite{12} where no fewer than 97 metrics used in the field of biomedicine alone are mentioned). However even some popular metrics have obvious drawbacks. For example the so-called $F_1$ metric,
\begin{equation}
    \begin{array}{ll}
    F_1=\dfrac{2TP}{2TP+FP+FN}     
    \end{array}
\end{equation}
is not symmetric in $P$ and $N$. Consequently if we swap the two classes then a value of $F_1$ close to $1$ can become a rather poor value close to $0$. Another metric commonly used in applications is the Accuracy defined as follows:
\begin{equation}     \begin{array}{ll}   
\text{Accuracy}=\dfrac{TP+TN}{TP+TN+FP+FN} \label{accu}
\end{array} \end{equation}
The problem of (\ref{accu}) is that it is not an informative metric when the two classes have significantly different sizes. In such a case we can trivially assign every test data sample to the larger class and achieve an Accuracy close to $1$.
\begin{figure}
    \centering
    \begin{tabular}{|c|c|c|c|}\cline{3-4}
         \multicolumn{2}{c|}{}&\multicolumn{2}{c|}{Classified/Predicted}\\\cline{2-4}
         \multicolumn{1}{c|}{}&Classes&Positive (1) &Negative (0)\\\cline{1-4}
         \multirow{2}{*}{\rotatebox[origin=c]{90}{\footnotesize{Actual}}}&Positive (1)& $C_{11}=TP$&$C_{12}=FN$\\\cline{2-4}
         &Negative (0)& $C_{21}=FP$&$C_{22}=TN$\\\hline
    \end{tabular}
    \caption{The confusion matrix for binary classification ($TP$ = true positive, $TN$= true negative, $FN$ = false negative, and $FP$ = false positive).}
    \label{tab:my_label}
\end{figure}

Several published comparison studies (e.g., see the papers cited above especially \cite{chicco},\cite{chicco2} and \cite{jurman}) have found that for binary classification the MCC metric, although not necessarily the best choice in all cases (see e.g. \cite{12}, \cite{13},\cite{mccref}), often yields a more informative and thus more useful performance score than many of the other metrics. In the next section we discuss the use of MCC for binary classification tasks. The main purpose of Section II is to introduce the basic concepts and emphasize a technical aspect of the MCC which is less discussed in the literature but is important as it confirms that MCC is a proper metric. Then in Section III we turn our attention to multinary (or multi-class) classification, which is the main topic of this paper, and begin with a discussion on the $\text{R}_{\text{K}}$ metric introduced in \cite{gorodkin}. In the same section we show that using the multivariate Pearson correlation (MPC) coefficients leads to two other metrics, and in Section IV, V and VII we introduce several new metrics for multinary classification.

\section{Binary Classification}
Consider an experiment of size $N$ (the total number of ``test data samples'' or ``cases'' that were analyzed and classified), which has produced the confusion matrix
\begin{equation}     \begin{array}{ll}     
\mathbf{C}=\begin{bmatrix} C_{11}&C_{12}\\C_{21}&C_{22}\end{bmatrix}
\end{array} \end{equation}
We assign the following numerical values to the two classes (see Fig. 1):
\begin{equation}     \begin{array}{ll}    
\text{Positive}=1 \\
\text{Negative}=0 \\ \label{eq:4}
\end{array} \end{equation}
and let the two binary sequences $\{t_n\}_{n=1}^N$ and $\{c_n\}_{n=1}^N$ indicate the true classes and, respectively, the classes assigned by the classifier. The Pearson correlation coefficient (PCC) of these two sequences has the following expression:
\begin{equation}     
\begin{array}{ll}  
 r&=\dfrac{\frac{1}{N}\sum\limits_{n=1}^N\left(t_n-\bar{t}\right)\left(c_n-\bar{c}\right)}{ \left[\frac{1}{N}\sum\limits_{n=1}^N \left(t_n-\bar{t}\right)^2 \frac{1}{N}\sum\limits_{n=1}^N \left(c_n-\bar{c}\right)^2\right]^{1 / 2} }\\
 &=\dfrac{\sum\limits_{n=1}^N t_n c_n-N \bar{t} \bar{c}} {
\left[\sum\limits_{n=1}^Nt_n^2-N \bar{t}^2\right]^{1 / 2}\left[\sum\limits_{n=1}^Nc_n^2-N \bar{c}^2\right]^{1 / 2} } \label{eq:5}
\end{array} 
\end{equation}
where
\begin{equation}     \begin{array}{ll}     
\bar{t}=\frac{1}{N}\sum\limits_{n=1}^N t_n ,\;\; \bar{c}=\frac{1}{N}\sum\limits_{n=1}^N c_n
\end{array} \end{equation}
are the means. Using the confusion matrix $\mathbf{C}$ and the values in (\ref{eq:4}) assigned to the two classes, we can evaluate all the quantities in (\ref{eq:5}) via relatively simple arguments. However, as a preparation for the more complicated case of multinary classification, we present below formulas for these quantities. Let
\begin{equation}     \begin{array}{ll}     
\mathbf{u}=[1, 1]^T\\\mathbf{v}=[1, 0]^T
\end{array} \end{equation}
Then using the fact that $t_n^2=t_n$ and $c^2_n=c_n$ we can easily verify that ($K=2$):
\begin{align}
 & N=\mathbf{u}^T\mathbf{C}\mathbf{u} = \sum\limits_{k=1}^K \sum\limits_{l=1}^K C_{kl}=TP+TN+FP+FN \label{eq:8}\\
&\sum\limits_{n=1}^N t_n^2=\sum\limits_{n=1}^N t_n=\bv^T\bC\bu =  \sum\limits_{l=1}^K C_{1l}=TP+FN\\
&\sum\limits_{n=1}^N c_n^2=\sum\limits_{n=1}^N c_n=\bu^T\bC\bv =  \sum\limits_{k=1}^K C_{k1}=TP+FP\\
&\sum\limits_{n=1}^N t_n c_n=\bv^T\bC\bv =  C_{11}=TP \label{eq:11} \end{align}  
Inserting the above expressions in (\ref{eq:5}) yields
\begin{equation}
r=\frac{\frac{TP}{N}-\frac{T P+F N}{N} \frac{T P+F P}{N}}{\left[\frac{T P+F N}{N} \frac{T P+F P}{N}\left(1-\frac{T P+F N}{N}\right)\left(1-\frac{T P+F P}{N}\right)\right]^{1 / 2}} \label{eq:12}
\end{equation}
which, after using the expression for $N$ in (\ref{eq:8}), becomes:
\begin{equation}
\text{MCC}=\frac{T P \cdot T N-F P \cdot F N}{[(T P+F N)(T P+F P)(T N+F P)(T N+F N)]^{1 / 2}} \label{eq:13}
\end{equation}
Equation (\ref{eq:12}) is the original formula for MCC that appeared in \cite{mat} and (\ref{eq:13}) is a slightly simplified (and possibly more intuitive) expression commonly used in the more recent literature (e.g. \cite{chicco} \cite{grand} \cite{chicco2}). We note in passing that in the statistical literature the MCC is known as Pearson phi coefficient (which is the PCC for two binary variables and has the same expression as MCC see e.g. \cite{cramer}).

An interesting aspect, which is not discussed in the literature as often as it should, concerns the values 1 and 0 assigned to the two classes. This choice may seem natural but it is arbitrary and this observation begs the question whether choosing other values in lieu of 1 and 0 would change the expression of MCC in (\ref{eq:13}). To answer this question we note that the PCC is invariant to the translation and scaling of the two sequences. In other words, if we replace $\{t_n\}$ and $\{c_n\}$ with
\begin{equation}
\theta_n=a+b t_n \text { and } \gamma_n=\alpha+\beta c_n\; (b, \beta>0)
\end{equation}
then the PCC of $\{\theta_n\}$ and $\{\gamma_n\}$ is identical to that of $\{t_n\}$ and $\{c_n\}$ (this is easy to see: the translation terms ($a,\alpha$) are absorbed in the means of the two sequences, therefore $(\theta_n-\bar{\theta})=b(t_n-\bar{t}),(\gamma_n-\bar{\gamma})=\beta(c_n-\bar{c})$, and the scaling factors ($b,\;\beta$) cancel in the fraction in (\ref{eq:5})). The consequence of this fact is that we can replace the values $1$ and $0$ assigned to the two classes by any other numbers and the MCC will not be affected. This is a desirable property without which MCC would not be a proper metric.

\section{Multinary classification: the $\text{R}_{\text{K}}$ and MPC metrics}
In this section we consider the general case of $K$ classes (with $K \geq 2$). Somewhat similarly to the discussion in the previous section (for $K=2$) we introduce two sequences for each class:
\begin{equation}
\left\{t_n(k)\right\}_{n=1}^N \text { and }\left\{c_n(k)\right\}_{n=1}^N
\end{equation}
where $t_n(k)=1$ if the $n-$th case is in the $k-$th class, and $=0$ otherwise; also $c_n(k)=1$ if the $n-$th case was assigned by the classifier to the $k-$th class, else $c_n(k)=0$. Let
\begin{equation}
\mathbf{t}_n=\left[t_n(1), \cdots, t_n(K)\right]^{\top}, \mathbf{c}_n=\left[c_n(1), \cdots, c_n(K)\right]^{\top}
\end{equation}
After this notational preamble, we move on to discuss the $\text{R}_{\text{K}}$ metric proposed in \cite{gorodkin}. In the second part of this section we make use of the MPC to obtain two other metrics.
\subsection{The $\text{R}_{\text{K}}$ metric}
The (sample) covariance matrix of the vectors $\mathbf{t}_n$ and $\mathbf{c}_n$ is given by:
\begin{equation}
\mathbf{R}_{t c}= \frac{1}{N} \sum_{n=1}^N\left(\mathbf{t}_n-\bar{\mathbf{t}}\right)\left(\mathbf{c}_n-\bar{\mathbf{c}}\right)^{\top}
\end{equation}
where
\begin{equation}
\bar{\mathbf{t}}= \frac{1}{N}\sum_{n=1}^N \mathbf{t}_n,\;\bar{\mathbf{c}}=\frac{1}{N}\sum_{n=1}^N \mathbf{c}_n
\end{equation}
The $K\times K$ matrices $\mathbf{R}_{tt}$ and $\mathbf{R}_{cc}$ are similarly defined. Then the so-called extended correlation coefficient $\text{R}_{\text{K}}$ introduced in \cite{gorodkin} is defined as follows:
\begin{equation}
\begin{array}{ll}
\text{R}_{\text{K}}&=\frac{\operatorname{tr}\left(\mathbf{R}_{t c}\right)}{\left[\operatorname{tr}\left(\mathbf{R}_{t t}\right) \operatorname{tr}\left(\mathbf{R}_{c c}\right) \right]^{1 / 2}}\\
&=\dfrac{\sum\limits_{k=1}^K\left[{R}_{t c}\right]_{k k}}{\left[\sum \limits_{k=1}^K\left[{R}_{t t}\right]_{k k}\right]^{1 / 2}\left[\sum \limits_{k=1}^K \left[{R}_{c c}\right]_{k k}\right]^{1/2}}
\end{array}
\end{equation}
where
\begin{equation}
\left[{R}_{tc}\right]_{k k}= \frac{1}{N} \sum_{n=1}^N\left[t_n(k)-\bar{t}_k\right]\left[c_n(k)-\bar{c}_k\right] \label{eq:20}
\end{equation}
and similarly for $[R_{tt}]_{kk}$ and $[R_{cc}]_{kk}$. The numerator and denominator in the above formula can be expressed as functions of the elements of the confusion matrix $\mathbf{C}$. We defer a discussion on this aspect until the other two metrics are discussed in the next sub-section.
\subsection{The MPC metrics}
The MPC is a matrix the $(k,l)$ element of which is given by:
\begin{equation}
\frac{\left[R_{t c}\right]_{k l}}{\left(\left[R_{t t}\right]_{k k}\left[R_{c c}\right]_{ll}\right)^{1 / 2}} \label{eq:25}
\end{equation}
The normalized trace of this matrix can be used to obtain a scalar performance metric:
\begin{equation}
\text{MPC}_1=\frac{1}{K} \sum_{k=1}^{K} \frac{\left[R_{t c}\right]_{k k}}{\left(\left[R_{t t}\right]_{k k}\left[R_{c c}\right]_{k k}\right)^{1 / 2}} \label{eq:25a}
\end{equation}
MPC$_1$ is the average of the univariate PCC metrics associated with the $K$ classes and it is probably the most natural extension of the MCC to the case of more than two classes. 

Similar to the fact that there is no unique solution to the problem of compressing the information in the confusion matrix $\mathbf{C}$ in a single number, there is no widely accepted scalar metric for the information contained in the correlation matrix in (\ref{eq:25}). MPC$_1$ and $\text{R}_{\text{K}}$ are just two possible metrics. The next metric is another quite natural choice:
\begin{equation}
\text{MPC}_2=  \frac{\sum \limits_{k=1}^{K}\left[R_{t c}\right]_{k k}}{\sum \limits_{k=1}^{K}\left(\left[R_{t t}\right]_{k k}\left[R_{c c}\right]_{k k}\right)^{1 / 2}} \label{eq:25b}
\end{equation}
Like $\text{R}_{\text{K}}$ and MPC$_1$, the above metric also lies in the interval $[-1,1]$. Indeed, because:
\begin{equation}
\left|\left[R_{tc}\right]_{kk} \right| \leq \left(\left[R_{tt}\right]_{kk}\right)^{1/2} \left(\left[R_{cc}\right]_{kk}\right)^{1/2} 
\end{equation}
it follows that
\begin{equation}
|\text{MPC}_2| \leq  \frac{\sum \limits_{k=1}^{K}|\left[R_{t c}\right]_{k k}|}{\sum \limits_{k=1}^{K}\left(\left[R_{t t}\right]_{k k}\right)^{1/2} \left(\left[R_{c c}\right]_{k k}\right)^{1 / 2}} \leq 1
\end{equation}
which proves the assertion. Interestingly, we also have that 
\begin{equation}
|\text{R}_{\text{K}}| \leq |\text{MPC}_2| \label{conj}
\end{equation}
Therefore MPC$_2$ is always closer to $+1$ or $-1$ than $\text{R}_{\text{K}}$, which appears to be an advantage of MPC$_2$ because $\text{R}_{\text{K}}$ was sometimes found to be relatively far from $- 1$ in situations in which it should have been close. To prove (\ref{conj}) use the Cauchy-Schwartz inequality to verify that the denominators of MPC$_2$ and $\text{R}_{\text{K}}$ satisfy the following inequality:
\begin{equation}
\sum_{k=1}^{K}\left(\left[R_{t t}\right]_{k k} \right)^{1/2}\left(\left[R_{c c}\right]_{k k}\right)^{1/2} \leq \left(\sum_{k=1}^{K}\left[R_{t t}\right]_{k k} \right)^{1/2}\left(\sum_{k=1}^{K}\left[R_{c c}\right]_{k k}\right)^{1/2} 
\end{equation}
from which (\ref{conj}) follows. However, as we will show in Section VI, the $\text{R}_{\text{K}}$ and MPC$_2$ metrics typically take on quite similar (or even identical) values and thus the possible advantage of MPC$_2$ implied by (\ref{conj}) is not significant.

Next we note that the three metrics reduce to the MCC in the case of $K=2$. This is not immediately obvious because in this section we have used two pairs of sequences for $K=2$, while only one pair was used in the previous section. However the second pair used here is redundant. Indeed (for $n=1,\cdots,N$),
\begin{equation}
\begin{array}{ll}
t_n(2) & = 1- t_n(1) \\
c_n(2) & = 1-c_n(1) 
\end{array} \label{32}
\end{equation}
which implies that:
\begin{equation}
\begin{array}{ll}
t_n(2) - \bar{t}_2 & = - \left[t_n(1) - \bar{t}_1 \right] \\
c_n(2) - \bar{c}_2 & = -\left[c_n(1) - \bar{c}_1 \right]
\end{array}
\end{equation}
Therefore,
\begin{equation}
\left[R_{tc}\right]_{11} = \left[R_{tc}\right]_{22}, \;\left[R_{tt}\right]_{11} = \left[R_{tt}\right]_{22}, \; \left[R_{cc}\right]_{11} = \left[R_{cc}\right]_{22} \label{eq:24}
\end{equation}
Using (\ref{eq:24}) in the formulas (\ref{eq:25}), (\ref{eq:25a}) and (\ref{eq:25b}) for the three metrics under discussion we can readily verify that:
\begin{equation}
\text{R}_{\text{K}} = \text{MPC}_1 = \text{MPC}_2 = \text{MCC} \; (\text{for}\; K=2)
\end{equation}

What remains to be done is to evaluate the covariances appearing in the formulas for the above three metrics (see e.g. (\ref{eq:20})) using the confusion matrix $\mathbf{C}=[C_{kp}]_{k,p=1}^K$. A calculation similar to (\ref{eq:8})-(\ref{eq:11}) yields the expressions:
\begin{equation}
\begin{aligned}
& N= \sum_{k=1}^K \sum_{p=1}^K C_{k p} \\
& \sum_{n=1}^N t_n^2(k)=\sum_{n=1}^N t_n(k)=\sum_{p=1}^K C_{k p} \triangleq \alpha_k \\
& \sum_{n=1}^N c_n^2(k)=\sum_{n=1}^N c_n(k)=\sum_{p=1}^K C_{p k} \triangleq \beta_k \\
& \sum_{n=1}^N t_n(k) c_n(k)=C_{k k} \label{quant}
\end{aligned}
\end{equation}
from which it follows that:
\begin{equation}\label{37}
\text{R}_{\text{K}}=\frac{\sum\limits_{k=1}^K\left[N C_{k k}-\alpha_k \beta_k\right]}{\left[\sum \limits_{k=1}^K \alpha_k\left(N-\alpha_k\right) \right]^{1/2} \left[\sum \limits_{k=1}^K \beta_k\left(N-\beta_k\right)\right]^{1 / 2}} 
\end{equation}
\begin{equation}
\text{MPC}_1 =\frac{1}{K}\sum_{k=1}^K\frac{\left[N C_{k k}-\alpha_k \beta_k\right]}{\left[\alpha_k \beta_k\left(N-\alpha_k\right)  \left(N-\beta_k\right)\right]^{1 / 2}} \label{38}
\end{equation}
\begin{equation}
\text{MPC}_2 =\frac{\sum \limits_{k=1}^K\left[N C_{k k}-\alpha_k \beta_k\right]}{\sum \limits_{k=1}^K\left[\alpha_k \left(N-\alpha_k\right)  \beta_k\left(N-\beta_k\right)\right]^{1 / 2}} \label{39}
\end{equation}
Finally note that instead of the simple average in MPC$_1$ (and possibly in the other two metrics as well) we can use a weighted average, for instance giving more weight to the smaller classes. 

\section{Multinary classification: the enhanced $\text{R}_{\text{K}}$ ($\text{ER}_{\text{K}}$) and MPC (EMPC) metrics}
 All the three metrics discussed in the previous section are equal to one in the case of perfect classification:
\begin{equation}
\begin{array}{ll}
\alpha_k = \beta_k = C_{kk} \; (k = 1,\cdots,K) &\implies  \\
&\text{R}_{\text{K}} = \text{MPC}_1 = \text{MPC}_2 = 1 \label{int}
\end{array}
\end{equation}
which can be readily seen from the formulas in (\ref{37})-(\ref{39}). However at the other extreme of fully unsatisfactory classification (when the confusion matrix is hollow, i.e. $C_{kk}=0$ for $k=1,\cdots,K$) it is known that $\text{R}_{\text{K}}$ can be rather far from $-1$ and therefore it does not clearly indicate the poor performance of the classifier in such a case. The MPC metrics, despite the result in (\ref{conj}) that shows that at least MPC$_2$ is closer to $-1$ than $\text{R}_{\text{K}}$, suffer from the same problem (see the numerical study in Section VI). The goal of this section is to present enhanced versions of these three metrics that do not suffer from this problem. 

Interestingly it turns our that the problem lies in the definition of the sequences $\{t_n(k), c_n(k)\}$ associated with the $k-$th class. Indeed, even in the case in which these sequences have no common ones (hence $C_{kk} = 0$) they still have lots of zeros in common and therefore their PCC can be rather far from $-1$. The solution is to eliminate the common zeros, which can be done by reducing the dimension of these sequences $\{t_n(k), c_n(k)\}_{n=1}^N$. The minimum dimension of these sequences, which allows the inclusion of all their ones, is $\alpha_k+\beta_k-C_{kk}$. The expressions for the three metrics corresponding to the reduced-dimension sequences are the same with one exception: $N$ in (\ref{37})-(\ref{39}) must be replaced by $(\alpha_k+\beta_k-C_{kk})$, after reinstating the factor $\frac{1}{N^2}$ that was cancelled in (\ref{37})-(\ref{39}) (doing the same in (\ref{38}) is not needed as it would not change anything). However doing so leads to an undefined case in the situation of perfect classification (\ref{int}). To see this consider, for example, the $k-$th term of MPC$_1$ (modified as indicated above):
\begin{equation}
\dfrac{(\alpha_k+\beta_k-C_{kk}) C_{kk} - \alpha_k \beta_k}{\left[\alpha_k \beta_k (\beta_k-C_{kk})(\alpha_k-C_{kk}) \right]^{1/2}} = \frac{0}{0} \; \text{for}\; \alpha_k = \beta_k = C_{kk}. \label{int1}
\end{equation}
Consequently, it is preferable to use a larger dimension than $\alpha_k+\beta_k -C_{kk}$. For the sake of simplicity we will use $\alpha_k+\beta_k$ (assuming that $\alpha_k+\beta_k<N$, which is true for most confusion matrices with the possible exception of some extremely imbalanced ones). The corresponding expressions for the three enhanced metrics are as follows:
\begin{equation}
\text{ER}_{\text{K}}=\dfrac{\sum \limits_{k=1}^K \frac{C_{k k}}{\alpha_k + \beta_k}}{\sum \limits_{k=1}^K \frac{\alpha_k \beta_k} {\left(\alpha_k +\beta_k\right)^{2}}} -1 \label{e1}
\end{equation}
\begin{equation}
\text{EMPC}_1= \frac{1}{K} \sum \limits_{k=1}^K \dfrac{(\alpha_k + \beta_k)C_{kk}}{\alpha_k\beta_k} -1 \label{e2}
\end{equation}
\begin{equation}
\text{EMPC}_2= \text{ER}_{\text{K}} \label{eq:444}
\end{equation}
Observe that EMPC$_2$ coincides with $\text{ER}_{\text{K}}$. Also note that in the perfect classification case (see (\ref{int})) all the three enhanced metrics are equal to one, whereas in the case of a hollow confusion matrix they are equal to $-1$ (as desired). Finally we remark on the fact that the above metrics do not reduce to MCC for $K=2$. The intuitive reason is that, while the equality in (\ref{32}) still holds, the dimensions of the two pairs of sequences in the mentioned equation are different. Nevertheless our experience is that the differences between MCC and any of the above metrics (with $K=2$) are quite small.  
\section{Multinary classification: the extended MCC (EMCC) metric}
In this section we introduce another new metric, which we call extended MCC (EMCC) because it can be viewed as an extension of the MCC to the multiclass case (as explained below). The EMCC has the following expression
\begin{equation}
\text{EMCC}=\frac{\prod \limits_{k=1}^K C_{k k}-\left[\prod \limits_{k=1}^K\left(\alpha_k-C_{k k}\right)\left(\beta_k-C_{k k}\right)\right]^{1 / 2}}{\left[\prod \limits_{k=1}^K \alpha_k \beta_k\right]^{1 / 2}} \label{EMCC}
\end{equation}
where $\{\alpha_k\}$ and $\{\beta_k\}$ are as defined in (\ref{quant}). Like the other metrics discussed in Section III and IV, EMCC also takes values in the interval $[-1,1]$. Specifically if $\mathbf{C}$ is a diagonal matrix then $\alpha_k=\beta_k=C_{kk}$ and therefore $\text{EMCC} =1$ (by a limit argument this also holds if some $C_{kk}=0$). In the other extreme case in which all $C_{kk}=0$ (for $k=1,\dots,K$), we get $\text{EMCC} =-1$ (once again by a limit argument one can verify that this is also true if some $\alpha_k\beta_k=0$). Furthermore in general we have that:
\begin{equation}
\text { EMCC } \leqslant \frac{\prod \limits_{k=1}^K C_{k k}}{\left[\prod \limits_{k=1}^K \alpha_k \beta_k\right]^{1 / 2}} \leqslant \frac{\prod \limits_{k=1}^K C_{k k}}{\left[\prod \limits_{k=1}^K C_{kk}^2\right]^{1 / 2}} = 1
\end{equation}
and
\begin{equation}
\text{EMCC} \geqslant-\frac{\left[\prod \limits_{k=1}^K\left(\alpha_k-C_{k k}\right)\left(\beta_k-C_{k k}\right)\right]^{1/2}}{\left[\prod \limits_{k=1}^K \alpha_k \beta_k\right]^{1 / 2}} \geqslant-1
\end{equation}
which proves that $\text{EMCC}\in [-1,1]$. Finally to show that EMCC reduces to MCC for $K=2$ observe that in the two-class case:
\begin{equation}
\begin{array}{ll}
\text{EMCC} & =\frac{C_{11} C_{22}-\left(C_{12} C_{21} C_{21} C_{12}\right)^{1 / 2}}{\left[\left(C_{11}+C_{12}\right)\left(C_{11}+C_{21}\right)\left(C_{22}+C_{21}\right)\left(C_{22}+C_{12}\right)\right]^{1 / 2}} \\
& =\frac{T P \cdot T N-F P \cdot F N}{[(T P+F N)(T P+F P)(T N+F P)(T N+F N)]^{1 / 2}} \\
& = \text{MCC}
\end{array}
\end{equation}
The intuition behind the EMCC formulation can be briefly explained as follows. The numerator in (\ref{EMCC}) is a ``measure'' of the difference between the diagonal elements of $\mathbf{C}$ and the off-diagonal elements (the more diagonal the matrix $\mathbf{C}$ is the larger the numerator, and vice versa). In (\ref{EMCC}) the product is preferred to summation because the former is much more sensitive to imbalances than the latter: for example if $C_{11}=0$ then first term in the numerator of (\ref{EMCC}) is equal to zero thus identifying the classification result as relatively poor, which would not happen if summation was used. The use of the square-root in (\ref{EMCC}) is to make the EMCC invariant to the scaling of the elements of $\mathbf{C}$ (if $\mathbf{C}$ is replaced by $\mu\mathbf{C}$, any $\mu>0$, then the EMCC is not effected). Finally the denominator in (\ref{EMCC}) is a ``normalization'' (which makes EMCC lie in the interval $[-1,1]$).

The EMCC can also at least partly be motivated in terms of correlations. To do so write (\ref{EMCC}) in the following form (making use of (\ref{quant})):
\begin{equation}\label{100}
\begin{array}{ll}
\text{EMCC}=\prod \limits_{k=1}^K \frac{\sum \limits_{n=1}^N t_n(k) c_n(k)}{\left[\sum \limits_{n=1}^N t_n^2(k) \sum \limits_{n=1}^N c_n^2(k)\right]^{1 / 2}} \\
\quad \quad \quad -\prod \limits_{k=1}^K\left[\frac{\sum \limits_{n=1}^N t_n(k)\left[t_n(k)-c_n(k)\right] \sum \limits_{n=1}^N c_n(k)\left[c_n(k)-t_n(k)\right]}{\sum \limits_{n=1}^N t_n^2(k) \sum \limits_{n=1}^N c_n^2(k)}\right]^{1/2}
\end{array}
\end{equation}

The generic factor of the first product in \eqref{100} measures the similarity (or correlation) of $\{t_n(k)\}$ and $\{c_n(k)\}$: this factor is equal to $1$ when $\{t_n(k)\}=\{c_n(k)\}(\forall n)$ and decreases towards $0$ as the two sequences overlap less and less. The generic factor of the second product in \eqref{100} also measures the closeness (or correlation) of $\{t_n(k)\}$ to $\{c_n(k)\}$  but in reverse order: it equals 0 when $\{t_n(k)\}=\{c_n(k)\}(\forall n)$ and increases towards 1 as the sequences become more and more dissimilar. Regarding the use of the product in \eqref{100} (instead of summation), this has already been motivated. Note that either term in \eqref{100} could in principle be used as a metric in its own right (and with a rather clear multivariate correlation flavor). However, because the two terms strengthen the effect on each other of the correlation (or lack thereof) between $\{t_n(k)\}\text{ and }\{c_n(k)\}$, their combination as in \eqref{100} appears to be a better metric.

To conclude this section we note that if any $\alpha_k$ or $\beta_k$ is equal to zero (in other words at least one column or row of $\mathbf{C}$ contains only zeros) then a direct use of the EMCC formula in (\ref{EMCC}) would result in an undefined case: $0/0$. In such a case a limit argument, similar to the one employed for MCC in \cite{chicco}, should be used to compute the value of EMCC. 

\section{Numerical comparisons}
In this section we numerically compare the following metrics for multinary classification: $\text{R}_{\text{K}}$, MPC$_1$, MPC$_2$, $\text{ER}_{\text{K}}$, EMPC$_1$, EMPC$_2$, EMCC and the Accuracy (A) defined as follows:
\begin{equation}
\text{A}=2  \frac{\sum \limits_{k=1}^K C_{kk}}{N}-1
\end{equation}
All metrics take values in the interval $[-1,1]$. Of course no simulation study, which is limited by many factors, can determine the usefulness of a metric, only its practical use in many applications can. With this caveat in mind, we proceed to the description of the cases that will be considered in this section. Note that in all cases $K=5$ and $N=1000$. Also note that a Matlab code for the generation of confusion matrices with the structures below and the reproduction of all plots that follow can be downloaded from \url{https://in.mathworks.com/matlabcentral/fileexchange/128393-mpc}.



We consider the following types of confusion matrices (CM) that appear to be of practical interest:
\begin{itemize}
\item[1)] Diagonal CM ($C_{kl} = 0$ for $k\neq l$)
\item[2)] Diagonally dominant CM ($C_{kk}$ is (much) larger than $C_{kl}$ and $C_{pk}$ for $l,p \neq k$)
\item[3)] Hollow CM ($C_{kk}=0$ for $k=1,\cdots, K$)
\item[4)] Off-diagonally dominant CM ($C_{kk}$ is (much) smaller than $C_{kl}$
and $C_{pk}$ for $l,p \neq k$)
\item[5)] Nearly uniform CM ($\{C_{kl}\}$ have similar magnitudes for $k,l=1, \cdots, K$)
\item[6)] Imbalanced ($3,2$) CM (the first three classes are much larger than the last two: $C_{11} + C_{22}+C_{33} \approx 0.5 N$ and the rest of the elements of $\mathbf{C}$ have similar magnitudes).
\item[7)] Imbalanced ($1,4$) CM (the first class is much larger than the last four: $C_{11} \approx 0.9 N$, $C_{kk} = 0$ for $k=2,\cdots, K$, and the rest of the elements of $\mathbf{C}$ have similar magnitudes)

\end{itemize}

\begin{figure*}
\begin{centering}
\begin{tabular}{c}
\includegraphics[width=17cm,height=20cm]{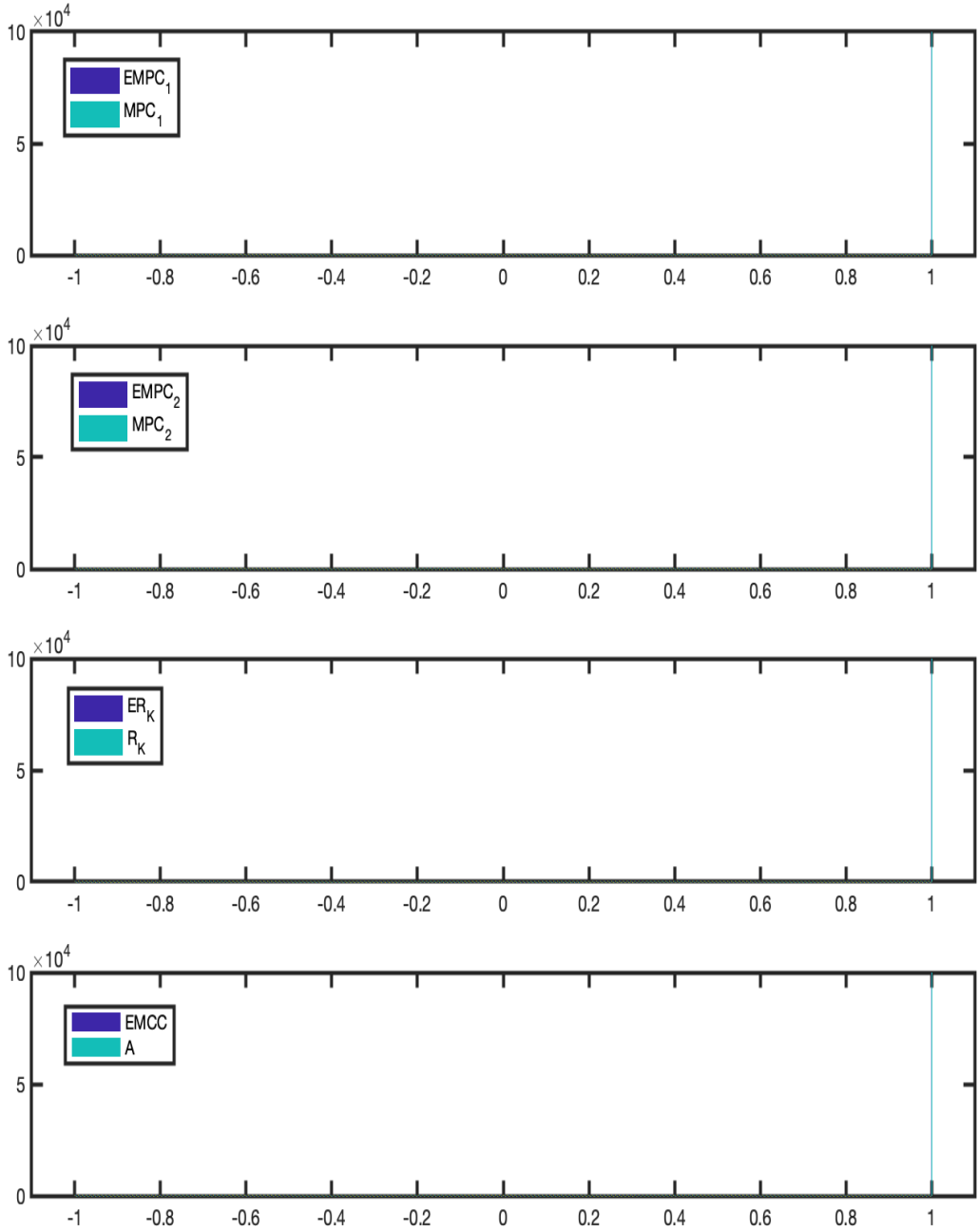} \tabularnewline
\tabularnewline
\end{tabular}
\par\end{centering}
\caption{Diagonal CM.}
\label{fig:1}
\end{figure*}
\begin{figure*}[ht]
\begin{centering}
\begin{tabular}{c}
\includegraphics[width=17cm,height=20cm]{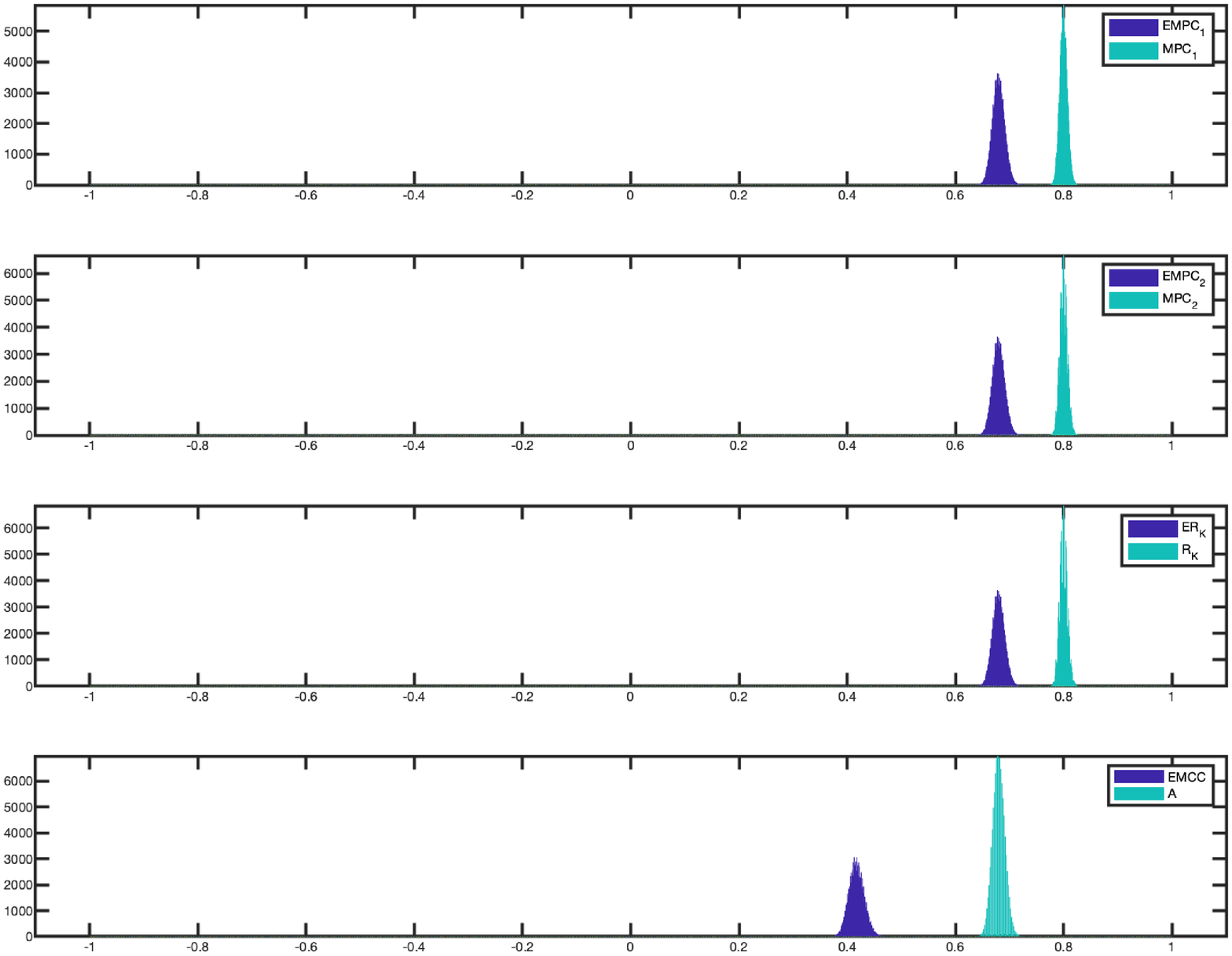} \tabularnewline
 \tabularnewline
\end{tabular}
\par\end{centering}
\caption{Diagonally dominant CM.}
\label{fig:2}
\end{figure*}

\begin{figure*}[ht]
\begin{centering}
\begin{tabular}{c}
\includegraphics[width=17cm,height=20cm]{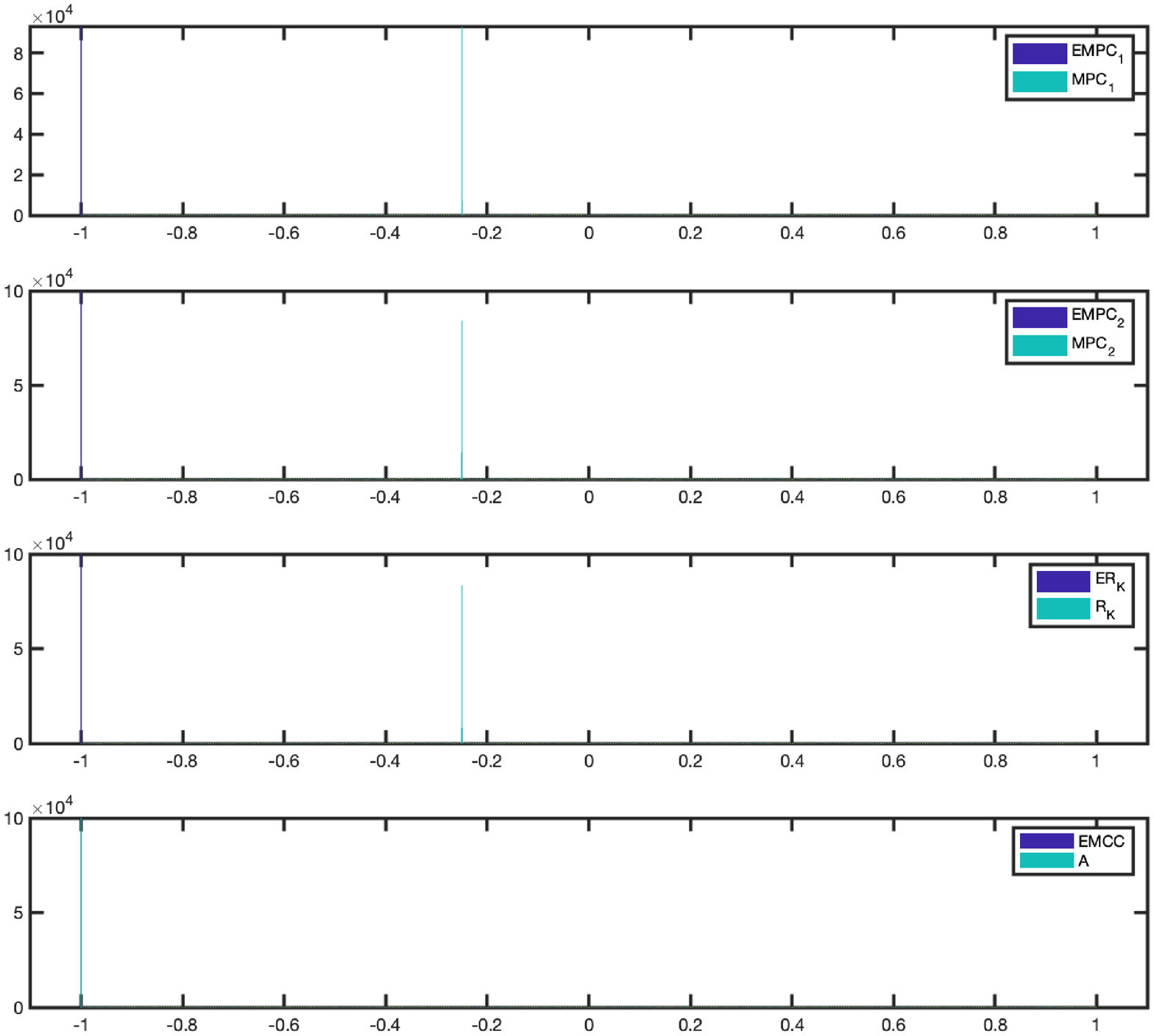} \tabularnewline
\tabularnewline
\end{tabular}
\par\end{centering}
\caption{Hollow CM.}
\label{fig:3}
\end{figure*}

\begin{figure*}[ht]
\begin{centering}
\begin{tabular}{c}
\includegraphics[width=17cm,height=20cm]{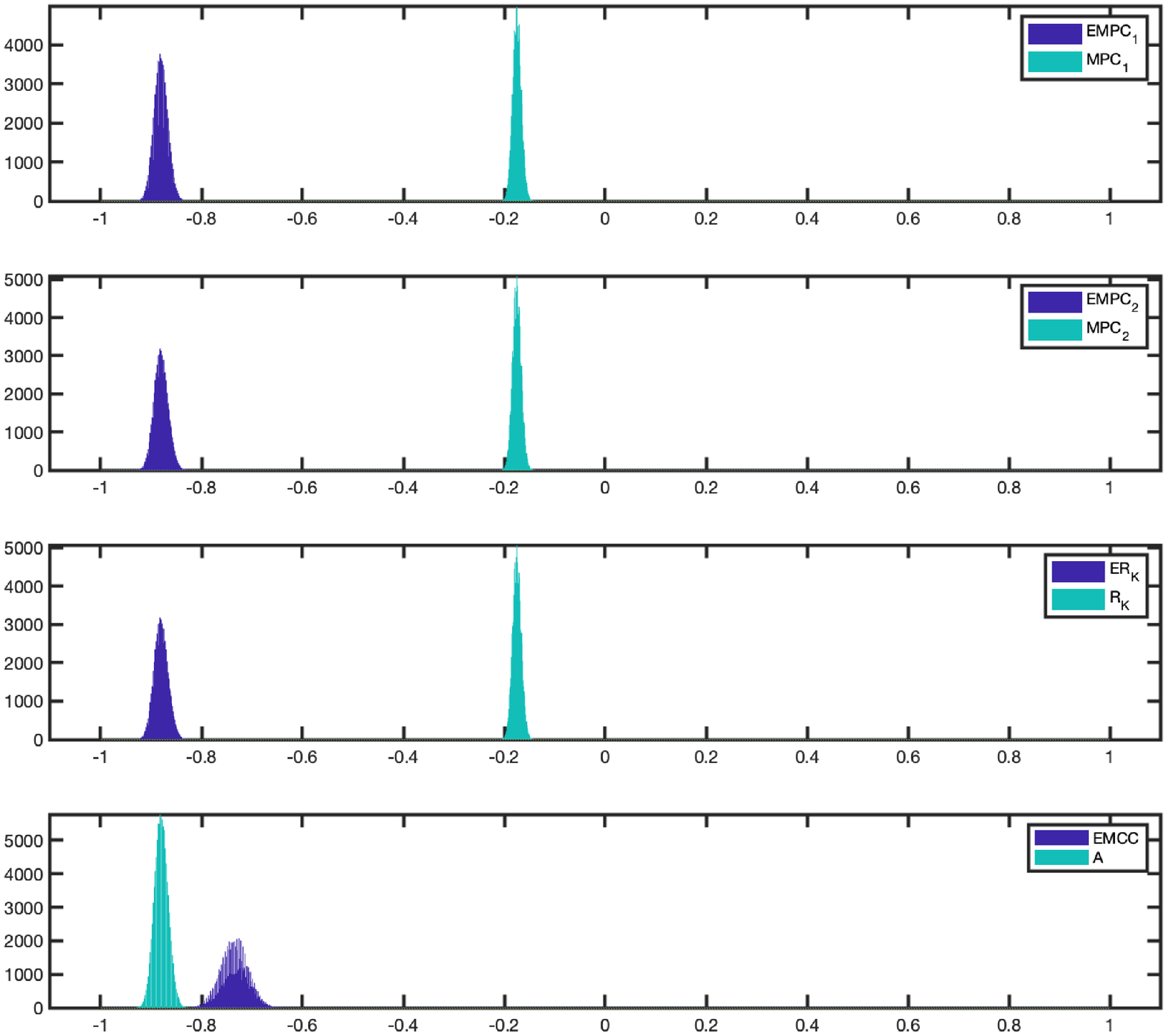} \tabularnewline
\tabularnewline
\end{tabular}
\par\end{centering}
\caption{Off-diagonally dominant CM.}
\label{fig:4}
\end{figure*}

\begin{figure*}[ht]
\begin{centering}
\begin{tabular}{c}
\includegraphics[width=17cm,height=20cm]{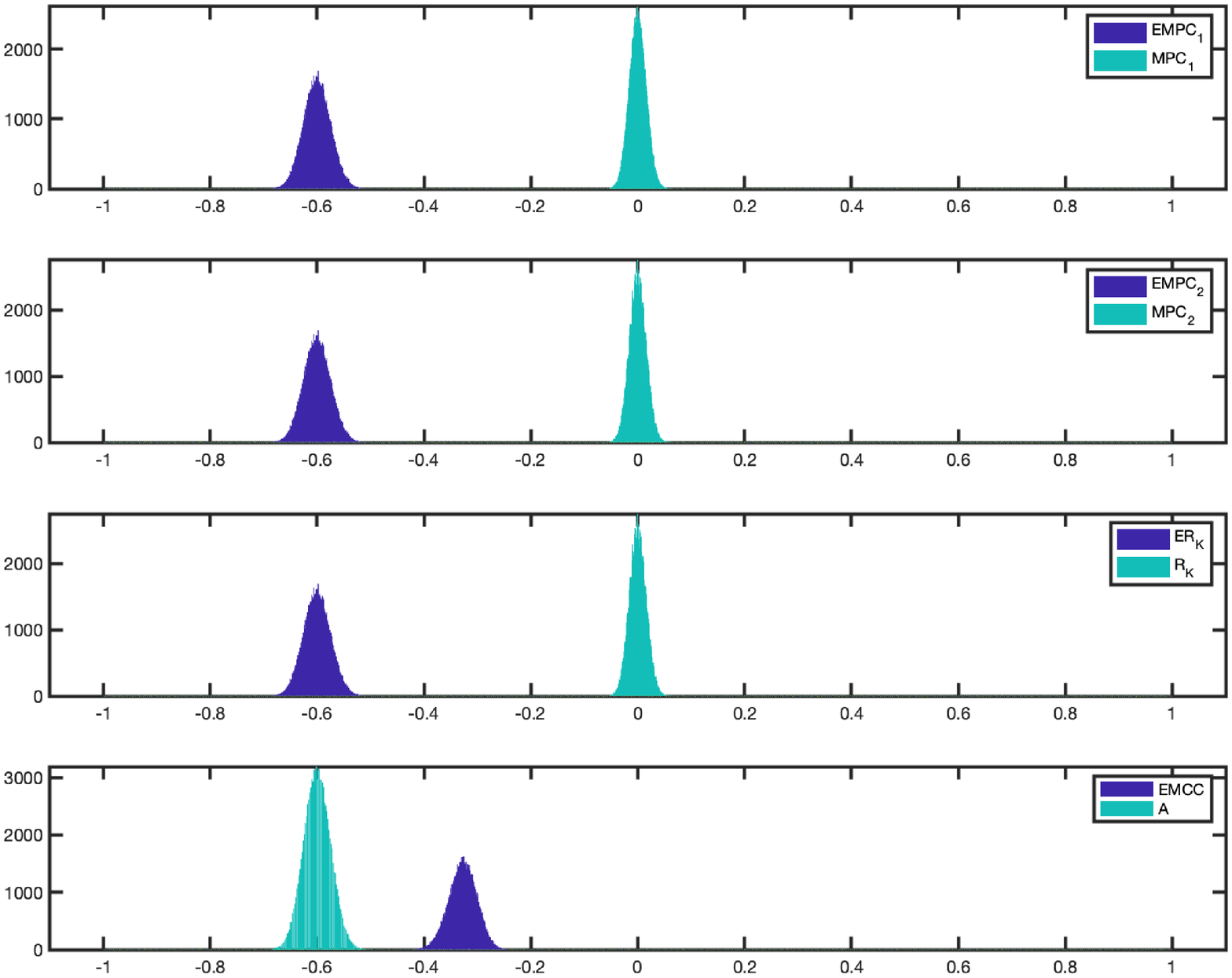} \tabularnewline
\tabularnewline
\end{tabular}
\par\end{centering}
\caption{Nearly uniform CM.}
\label{fig:5}
\end{figure*}

For each of the above structures we randomly generate $10^5$ CMs and compute the eight metrics. The histograms of the so-obtained metric values are shown in Figs. 1-7, which correspond to the seven cases of CM structures described above. We make the following remarks based on the results in these figures:
\begin{itemize}
\item[a)] For diagonal CMs all metrics are equal to one (as expected)
\item[b)] In most cases the $\text{R}_{\text{K}}$, MPC$_1$ and MPC$_2$ metrics take on quite similar values. The exception is the case of imbalanced CM in which MPC$_1$ appears to perform slightly better than the other two metrics. 
\item[c)] The enhanced metrics yield similar results in all cases (we remind the reader that $\text{ER}_{\text{K}}=$ EMPC$_2$, see (\ref{eq:444})). 
\item[d)] The $\text{ER}_{\text{K}}$, EMPC$_1$, and EMPC$_2$ metrics appear to always provide more informative results than $\text{R}_{\text{K}}$, MPC$_1$ and MPC$_2$. As predicted this is especially true in the cases of hollow or nearly hollow CMs but also for nearly uniform CMs or imbalanced CMs.
\item[e)] For imbalanced CMs, EMPC$_1$ and EMPC$_2$ = $\text{ER}_{\text{K}}$ appear to be the metrics of choice as they provide results that are more in concordance with the intuition than the other metrics.
\item[f)] In most cases the results obtained with A are quite similar to those provided by the enhanced metrics. The exception is the case of imbalanced CMs (the well-known Achilles heel of A) in which A yields rather counter-intuitive results, in contrast to the enhanced metrics that have negative values.
\item[g)] The enhanced metrics take on slightly more favorable values for the imbalanced ($1,4$) CM than for the imbalanced ($3,2$) CM. Because in the former case most of the test samples are correctly classified while in the latter the results are rather mixed with many samples being misclassified, the slight preference of the enhanced metrics for the former case appears justifiable (we note again that in both cases the enhanced metrics evaluate the classification results as poor, unlike A which evaluates the results as good to excellent).
\item[h)] Finally, the results obtained with EMCC are not far from those provided by the enhanced metrics and thus they appear to be quite reasonable. 
\end{itemize}

\begin{figure*}[ht]
\begin{centering}
\begin{tabular}{c}
\includegraphics[width=17cm,height=20cm]{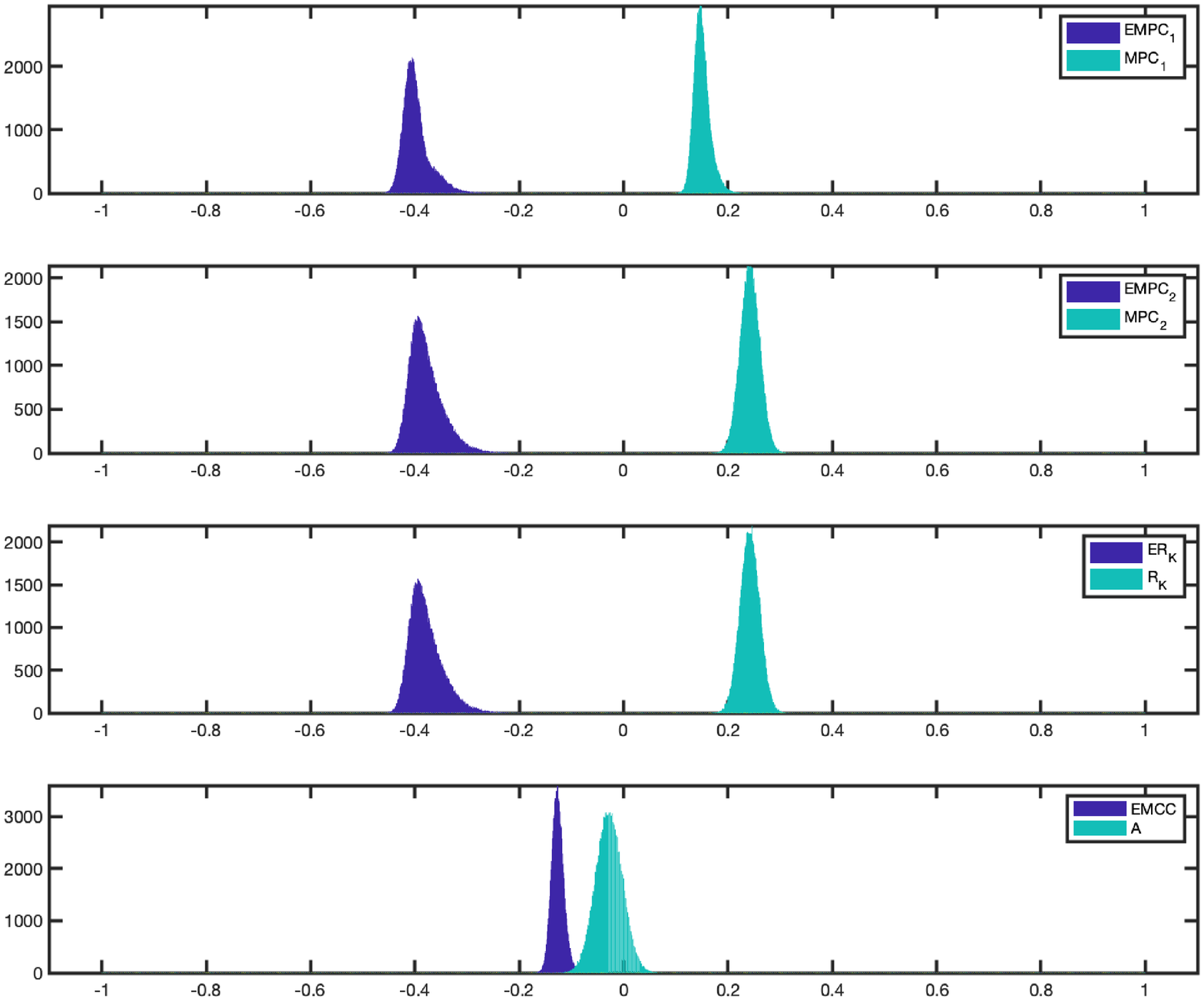} \tabularnewline
  
\tabularnewline
\end{tabular}
\par\end{centering}
\caption{Imbalanced ($3,2$) CM.}
\label{fig:6}
\end{figure*}

\begin{figure*}[ht]
\begin{centering}
\begin{tabular}{c}
\includegraphics[width=17cm,height=20cm]{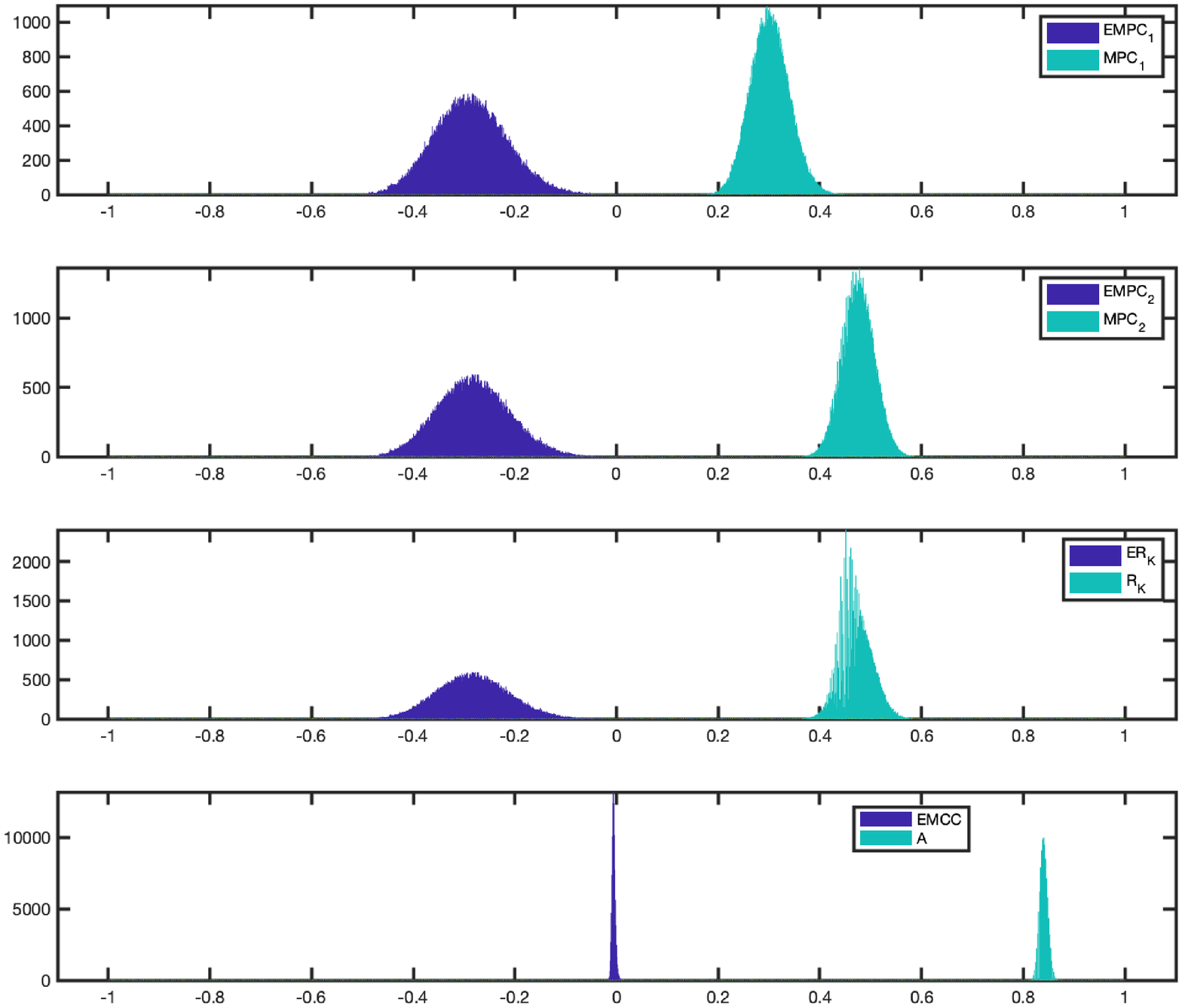} \tabularnewline
 \tabularnewline
\end{tabular}
\par\end{centering}
\caption{Imbalanced ($1,4$) CM.}
\label{fig:7}
\end{figure*}

\section{Future research}
We end the paper with a number of comments on a topic left for further research. As explained in Section IV, for the sake of simplicity we used $(\alpha_k+\beta_k)$ as the reduced dimension of the sequences $\{t_n(k)\}$ and $\{c_n(k)\}$ (indeed the enhanced metrics corresponding to this choice have quite simple expressions, see equations (\ref{37})-(\ref{38})). However this choice of the reduced dimension also means that the so-obtained sequences share $C_{kk}$ extraneous zeros which can affect the performance. To explain the effect of these zeros on the metrics consider a case in which their number is much larger than $\alpha_k$ or $\beta_k$ (such as when $N \gg 1$ and we use sequences of dimension $N$ as in Section IV). Then we have $\bar{t}_k \approx 0$ and $\bar{c}_k \approx 0$ and this implies that a pair $(t_n(k)=0, c_n(k)=1)$ (or vice-versa) does not contribute almost anything to the correlation between the sequences. The consequence is that poor classification performance is not detected (good performance will be detected because a pair $(t_n(k)=1, c_n(k)=1)$ increases the correlation).

The implication of the above discussion is that reducing the number of extraneous zeros is potentially a good idea. To implement this idea consider the following reduced dimension:
\begin{equation}
N_k = \alpha_k + \beta_k - \rho C_{kk}
\end{equation}
where $\rho =1$ corresponds to the minimum possible dimension and $\rho=0$ to the reduced dimension used in Section IV (as explained there we should avoid $\rho=1$, see (\ref{int1}), so we assume $\rho<1$). Like in Section IV consider the generic term of MPC$_1$ after reducing the dimension: 
\begin{equation}
\Delta_k= \frac{(\alpha_k + \beta_k - \rho C_{kk})C_{kk} - \alpha_k \beta_k}{\left[\alpha_k \beta_k (\alpha_k-\rho C_{kk})(\beta_k-\rho C_{kk})\right]^{1/2}} \quad (\rho<1)
\end{equation}
A straightforward calculation shows that:
\begin{equation}
\begin{array}{ll}
C_{kk} = 0 &\implies \Delta_k = 1   \\
\alpha_k = \beta_k = C_{kk}    &\implies \Delta_k = \frac{(1-\rho)C_{kk}^2}{(1-\rho)C_{kk}^2} = 1 \quad (\text{for}\; \rho<1) 
\end{array}
\end{equation}
Therefore any reduced dimension with $\rho<1$ leads to metrics that attain $+1$ for perfect classification and $-1$ in the case of complete misclassification. As explained above the number of extraneous zeros decreases as $\rho$ approaches one, so let us choose
\begin{equation}
\rho = 0.9
\end{equation}
The corresponding enhanced metrics have the following expressions:
\begin{equation}
\tilde{\text{E}}\text{R}_{\text{K}}=\dfrac{\sum \limits_{k=1}^K \left[N_k C_{k k}-\alpha_k \beta_k\right]/{N_k^2}}{\left[\sum \limits_{k=1}^K \alpha_k (\beta_k-\rho C_{kk})/N_k^2\right]^{1/2} \left[\sum \limits_{k=1}^K\beta_k (\alpha_k-\rho C_{kk})/N_k^2\right]^{1/2}}
\end{equation}
\begin{equation}
 \tilde{\text{E}}\text{MPC}_1=\dfrac{1}{K}\sum \limits_{k=1}^K\dfrac{N_k C_{kk}-\alpha_k \beta_k}{\left[\alpha_k \beta_k(\alpha_k-\rho C_{kk})(\beta_k-\rho C_{kk})\right]^{1/2}}   
\end{equation}
\begin{equation}
 \tilde{\text{E}}\text{MPC}_2=\dfrac{\sum \limits_{k=1}^K\left[N_k C_{kk}-\alpha_k \beta_k\right]/{N_k^2}}{\sum \limits_{k=1}^K{\left[\alpha_k \beta_k(\alpha_k-\rho C_{kk})(\beta_k-\rho C_{kk})\right]^{1/2}}/{N_k^2}}   
\end{equation}
While these expressions are a bit more complicated than the formulas corresponding to the choice $\rho=0$ (see (\ref{e1})-(\ref{eq:444})), the main question is how the former perform in comparison to the latter. The results of a preliminary comparison of $\text{EMPC}_1$, and $\tilde{\text{E}}\text{MPC}_1$, in the same seven cases as in the previous section, are presented in Fig. 9. As expected $\tilde{\text{E}}\text{MPC}_1$ penalizes misclassification more than $\text{EMPC}_1$, but more research will be needed to determine if the $\tilde{\text{E}}$ metrics are better choices than the simpler $\text{E}$ metrics in a particular application. 

Finally we remark on the fact that, like EMPCs, the $\tilde{\text{E}}\text{MPC}$ metrics do not reduce to MCC for $K=2$. However the latter metrics, unlike the former, can take on values that are quite different from the MCC values. To illustrate the possibly significant difference between $\tilde{\text{E}}\text{MPC}$ and MCC scores we consider the following imbalanced CM, 
\begin{equation}
\mathbf{C} = \begin{bmatrix}
    993 & 3 \\
    3 & 1
\end{bmatrix}
\end{equation}
for which:
\begin{equation}
\begin{array}{ll}
    \text{A}   &= 0.99  \\
    \text{MCC} &= 0.25  \; (=\text{MPC}_1 = \text{EMPC}_1) \\
    \tilde{\text{E}}\text{MPC}_1 &= -0.36 \; (\text{for}\; \rho \approx 1)
\end{array}
\end{equation}
As expected the A metric is biased toward the much bigger class and therefore considers the classification results to be almost perfect. The MCC, which is also somewhat biased toward the bigger class, finds these results to be above average. On the other hand the $\tilde{\text{E}}\text{MPC}_1$ scores them as poor. The application in hand will determine which evaluation is more in agreement with our intuition and expectations. In several practical applications the fact that $75\%$ of the test samples from the small class have been misclassified is important and consequently the $\tilde{\text{E}}\text{MPC}_1$ score appears to be more plausible/accurate. The possible implication is that the \textit{$\tilde{\text{E}}\text{MPC}$s may be the correlation-based metrics of choice not only for $K>2$ but even for $K=2$} - in the latter case they can provide more credible/precise scores than MCC especially in the case of imbalanced CMs (see \cite{13} for a critique of the MCC application to such CMs).
\begin{figure*}[ht]
\begin{centering}
\begin{tabular}{c}
\includegraphics[width=17cm,height=20cm]{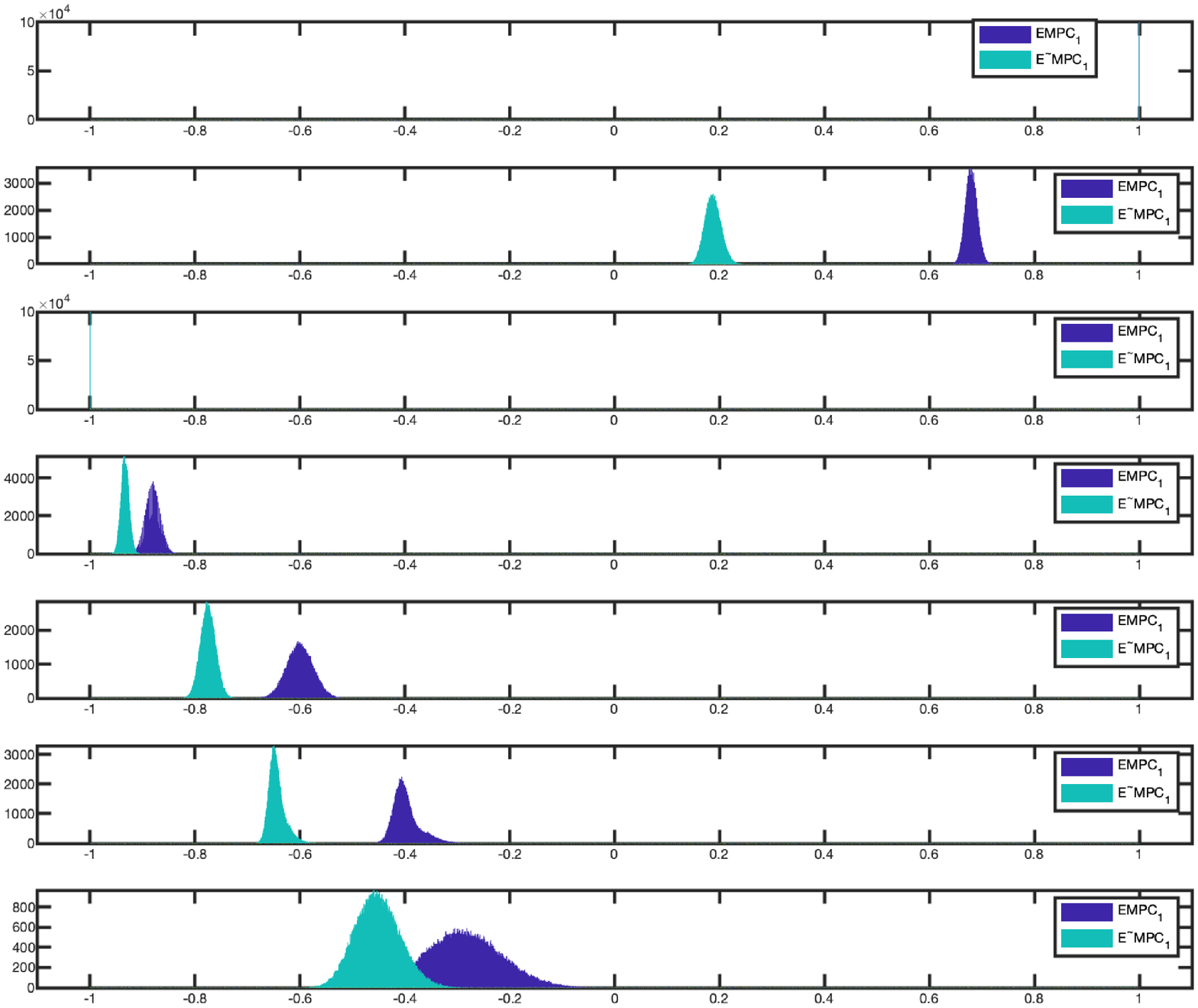} \tabularnewline
 \tabularnewline
\end{tabular}
\par\end{centering}
\caption{Comparison of EMPC$_1$ and $\tilde{\text{E}}\text{MPC}_1$ metrics in the seven cases of CM described in Section VI.}
\label{fig:8}
\end{figure*}

\section{Conclusions}
After an introductory discussion on the use of MCC for binary classification, we have focused on the main topic of the paper namely multinary (aka multiclass) classification. We have shown that besides $\text{R}_{\text{K}}$ there are two other metrics that extend the univariate Pearson correlation coefficient to the multivariate case, which we called $\text{MPC}_1$ and $\text{MPC}_2$ ($\text{MPC} =$ Multivariate Pearson Correlation). The latter metrics were found to behave quite similarly to $\text{R}_{\text{K}}$ and, more importantly, all three were found to provide unreliable (too optimistic) results in cases in which the classifier performed poorly. Motivated by this fact we have introduced  enhanced versions of them called $\text{ER}_\text{K}$, $\text{EMPC}_1$ and $\text{EMPC}_2$. We have shown that the enhanced metrics can yield results that are (much) more in agreement with the intuition than those obtained with $\text{R}_{\text{K}}$, $\text{MPC}_1$ and $\text{MPC}_2$, especially when the confusion matrix was hollow (or nearly hollow), uniform (or nearly so) or imbalanced. Additionally, the enhanced metrics have simple expressions and can be easily computed from the confusion matrix. The logical conclusion is that $\text{EMPC}_1$ or $\text{EMPC}_2=\text{ER}_{\text{K}}$ ($\text{ER}_{\text{K}}$ was shown to coincide with $\text{EMPC}_2$) should be used in preference to $\text{R}_{\text{K}}$, $\text{MPC}_1$ or $\text{MPC}_2$. Regarding choosing between $\text{EMPC}_1$ and $\text{EMPC}_2$, our experience is that these two metrics typically behave in quite a similar way, therefore either can be chosen for use in a practical application or better still the user can use both and compare the results. 

The numerical study of the paper has also found that the enhanced metrics and Accuracy often yield quite similar results except in the case of imbalanced confusion matrices in which the Accuracy results were unreliable (too optimistic). 

Finally, we have also introduced an additional new metric called $\text{EMCC}$ (Extended MCC) whose behaviour was shown to be quite satisfactory (not too far from that of the enhanced metrics). More numerical comparisons will be needed to determine its position in relation to the enhanced metrics and whether it can be preferred to the latter and thus be the method of choice in some applications. The same is true for the versions of the enhanced metrics discussed in the  section about future research. The latter metrics have a distinctive feature: by varying $\rho$ in their formulas the user can span the entire set of metrics from $\tilde{\text{E}}\text{MPC}$ (for $\rho \approx 1$) through  $\text{EMPC}$ (for $\rho = 0$) to $\text{MPC}$ and beyond  (for $\rho \ll -1$). Because the metrics penalize misclassification more and more as $\rho$ increases, the user has the option of choosing $\rho$ according to the application in hand. In our opinion this is an appealing feature that should be of interest especially to practitioners. Our current experience is that the $\tilde{\text{E}}\text{MPC}$ metrics have an advantage over the other correlation-based metrics (R$_{\text{K}}$, MPC, EMPC) for $K>2$ and even over MCC for $K=2$.

\bibliographystyle{ieeetr}
\bibliography{ref}

\begin{thebibliography}{10}

\bibitem{chicco}
D.~Chicco and G.~Jurman, ``{The advantages of the Matthews correlation
  coefficient (MCC) over F1 score and accuracy in binary classification
  evaluation},'' {\em BMC genomics}, vol.~21, pp.~1--13, 2020.

\bibitem{grand}
M.~Grandini, E.~Bagli, and G.~Visani, ``Metrics for multi-class classification:
  an overview,'' {\em arXiv preprint arXiv:2008.05756}, 2020.

\bibitem{ref6}
A.~Tharwat, ``Classification assessment methods,'' {\em Applied computing and
  informatics}, vol.~17, no.~1, pp.~168--192, 2021.

\bibitem{ref7}
M.~Sokolova and G.~Lapalme, ``A systematic analysis of performance measures for
  classification tasks,'' {\em Information processing \& management}, vol.~45,
  no.~4, pp.~427--437, 2009.

\bibitem{ref8}
M.~Hossin and M.~N. Sulaiman, ``A review on evaluation metrics for data
  classification evaluations,'' {\em International journal of data mining \&
  knowledge management process}, vol.~5, no.~2, p.~1, 2015.

\bibitem{ref9}
V.~Labatut and H.~Cherifi, ``Evaluation of performance measures for classifiers
  comparison,'' {\em arXiv preprint arXiv:1112.4133}, 2011.

\bibitem{chicco2}
D.~Chicco, M.~J. Warrens, and G.~Jurman, ``The matthews correlation coefficient
  (mcc) is more informative than cohen’s kappa and brier score in binary
  classification assessment,'' {\em IEEE Access}, vol.~9, pp.~78368--78381,
  2021.

\bibitem{jurman}
G.~Jurman, S.~Riccadonna, and C.~Furlanello, ``A comparison of mcc and cen
  error measures in multi-class prediction.,'' {\em Plos one}, vol.~7, no.~8,
  pp.~e41882--e41882, 2012.

\bibitem{12}
A.~Reinke, M.~D. Tizabi, M.~Baumgartner, M.~Eisenmann, D.~Heckmann-N{\"o}tzel,
  A.~E. Kavur, T.~R{\"a}dsch, C.~H. Sudre, L.~Acion, M.~Antonelli, {\em
  et~al.}, ``Understanding metric-related pitfalls in image analysis
  validation,'' {\em arXiv}, 2023.

\bibitem{13}
Q.~Zhu, ``On the performance of matthews correlation coefficient (mcc) for
  imbalanced dataset,'' {\em Pattern Recognition Letters}, vol.~136,
  pp.~71--80, 2020.

\bibitem{mccref}
A.~Kumar, A.~Niculescu-Mizil, K.~Kavukcoglu, and H.~Daum{\'e}, ``A binary
  classification framework for two-stage multiple kernel learning,'' in {\em
  Proceedings of the 29th International Coference on International Conference
  on Machine Learning}, pp.~1331--1338, 2012.

\bibitem{gorodkin}
J.~Gorodkin, ``{Comparing two K-category assignments by a K-category
  correlation coefficient},'' {\em Computational biology and chemistry},
  vol.~28, no.~5-6, pp.~367--374, 2004.

\bibitem{mat}
B.~W. Matthews, ``{Comparison of the predicted and observed secondary structure
  of T4 phage lysozyme},'' {\em Biochimica et Biophysica Acta (BBA)-Protein
  Structure}, vol.~405, no.~2, pp.~442--451, 1975.

\bibitem{cramer}
H.~Cram{\'e}r, {\em Mathematical methods of statistics}, vol.~26.
\newblock Princeton university press, 1999.

\end{thebibliography}

\end{document}